\documentclass[aps,prd,
groupedaddress,showpacs,nofootinbib,amssymb,final]{aipproc}

\layoutstyle{8x11single}

\usepackage[T1]{fontenc}
\usepackage[latin1]{inputenc}
\usepackage{graphicx}
\usepackage[english]{babel}
\usepackage{amsmath}
\usepackage{amssymb}
\usepackage{amsfonts}

\begin{document}

\def\pp{{\, \mid \hskip -1.5mm =}}
\def\cL{{\cal L}}
\def\be{\begin{equation}}
\def\ee{\end{equation}}
\def\bea{\begin{eqnarray}}
\def\eea{\end{eqnarray}}
\def\beq{\begin{eqnarray}}
\def\eeq{\end{eqnarray}}
\def\tr{{\rm tr}\, }
\def\nn{\nonumber \\}
\def\e{{\rm e}}

\title{Non-singular modified gravity: the unification of the inflation, dark
energy and dark mater}

\classification{95.36.+x, 98.80.Cq, 04.50.Kd, 11.10.Kk, 11.25.-w}

\keywords{dark energy, modified gravity}

\author{Shin'ichi Nojiri}{
address={Department of Physics, Nagoya University, Nagoya 464-8602, Japan}}

\author{Sergei D. Odintsov\footnote{
Also at CTP, Tomsk State Pedagogical Univ.}\ }
{address={Instituci\`{o} Catalana de Recerca i Estudis Avan\c{c}ats (ICREA), Barcelona \\ 
and \\
Institut de Ciencies de l'Espai (IEEC-CSIC),
Campus UAB, Facultat de Ciencies, Torre C5-Par-2a pl, \\
E-08193 Bellaterra (Barcelona), Spain}}

\begin{abstract}

Using the fluid representation, we formulate the conditions for the appearance 
of all four types finite-time future singularity in modified gravity 
in accelerating FRW universe. It stressed that number of standard quintessence/phantom 
DE theories (including scalar, fluid, DBI ones, etc)
brings the accelerating cosmology to future singularity precisely in the same 
way as singular modified gravity DE. The viable non-singular modified gravity unifying 
the early-time inflation with late-time acceleration is considered.
It is shown that adding such non-singular theory to another realistic singular 
modified gravity which has the accelerating solution with future singularity 
may cure the singularity of resulting combined model. This universal scenario 
may be naturally applied to standard singular DE models as well as to inflationary 
theories with initial singularity. This suggests the additional fundamental reasoning 
for modification of General Relativity. 

\end{abstract}

\maketitle

\section{Introduction}

Modified gravity suggests the gravitational alternative for the unified
description of the early-time inflation with late-time acceleration and
dark matter. It is remarkable that in this scenario for the universe
evolution there is no need in the introduction of extra dark components
like inflaton, quintessence, dark fluid, dark matter particles, etc.
The only early-time and late-time gravitational action is changed if compare with General
Relativity  in such a
way that local tests and cosmological bounds are not violated. Hence, the
gravitational field is responsible for dark behavior of the universe.
The comparison of different modified gravities aimed to the accelerating
universe description may be found in ref.\cite{Nojiri:2006ri}.

Imagine that we consider the (accelerating) FRW universe. In this case, 
whatever is modified gravity, it may be represented as General Relativity with some
(generalized) fluid. Different properties of
modified gravity may be clarified in this representation. For instance, for
realistic theories which unify
the early-time inflation with late-time acceleration and which we
consider here it is clear that they behave as effective
phantom/quintessence fluids at the late times. Then, such modified
gravities may bring the evolution to the finite-time future singularity of
one of four types according to the classification of
ref.\cite{Nojiri:2005sx} in the same way as
other phantom/quintessence dark energy models. There is no any qualitative
difference between modified gravity and other dark energy models in this
respect. (Reversing the time flow, the finite-time future singularity is
nothing else as initial singularity for number of inflationary theories.)

In the present paper we explicitly demonstrate that the singularity
appearance is typical property for number of modified gravity with the
effective quintessence/phantom late-time behavior precisely in the same
way as for other (scalar/fluid/composite) dark energy models. The appearance
of all four types of future singularity in unified modified gravity
occurs as it was demonstrated in ref.\cite{Nojiri:2008fk} (for first
observation on Big Rip singularity in modified gravity and its avoidance
by adding of $R^2$-term see ref.\cite{Abdalla:2004sw,Briscese:2006xu}).
The realistic non-singular modified gravity which unifies the early-time
inflation with
late-time acceleration may be considered as the tool to prevent the future
singularity. In particular, it is demonstrated that adding such
non-singular theory to another realistic modified gravity which has the
accelerating solution with future singularity may cure the singularity
appearance in combined
realistic modified gravity. In the same way, the initial singularity may 
be cured unless the classical description is valid.

\section{The formulation of modified gravity as General Relativity with generalized fluid
and finite-time future singularities}

Let us start from the general modified gravity with the action:
\be
\label{I}
S = \int d^4 x \sqrt{-g} \left\{ \frac{1}{2\kappa^2}
\left( R + f\left(R, R_{\mu\nu} R^{\mu\nu}, R_{\mu\nu\alpha\beta} R^{\mu\nu\alpha\beta},
\Box R, \Box^{-1} R, \cdots\right) \right) + L_\mathrm{m} \right\}\,,
\ee
where all combinations of local and non-local terms are possible, $L_\mathrm{m}$ is matter
Lagrangian and the function $f(R,\cdots)$ may also contain gravitational
 partners (say,
dilaton, axion, etc. in string-inspired gravity).

We consider spatially-flat FRW universe with scale factor $a(t)$.
In all cases for theory (\ref{I}), it is possible to write the gravitational
field equations in the form of standard FRW equations with effective energy-density
$\rho_\mathrm{eff}$ and pressure $p_\mathrm{eff}$ produced by the extra gravitational
terms $F(R,\cdots)$ and $L_\mathrm{m}$.

For instance, when $f=f(R)$, one gets
\bea
\label{Cr4}
\rho_\mathrm{eff} &=& \frac{1}{\kappa^2}\left(-\frac{1}{2}f(R) + 3\left(H^2 + \dot H\right) f'(R)
 - 18 \left(4H^2 \dot H + H \ddot H\right)f''(R)\right) + \rho_\mathrm{matter}\, ,\\
\label{Cr4b}
p_\mathrm{eff} &=& \frac{1}{\kappa^2}\left(\frac{1}{2}f(R) - \left(3H^2 + \dot H \right)f'(R)
+ 6 \left(8H^2 \dot H + 4{\dot H}^2
+ 6 H \ddot H + \dddot H \right)f''(R) + 36\left(4H\dot H + \ddot H\right)^2f'''(R) \right) \nn
&& + p_\mathrm{matter}\, .
\eea
For details of explicit fluid presentation of $f(R)$ gravity, see \cite{Capozziello:2005mj}.
In case of Gauss-Bonnet modified gravity $f(G)$ \cite{Nojiri:2005jg}:
\bea
\label{EoS3}
\rho_\mathrm{eff} &=& \frac{1}{2\kappa^2}\left[{\cal G} f_{\cal G}'({\cal G})
 - f_{\cal G}({\cal G}) - 24^2 H^4 \left(2 {\dot H}^2 + H\ddot H + 4H^2 \dot H\right) f_{\cal G}'' \right]
+ \rho_\mathrm{matter}\ ,\nn
p_\mathrm{eff} &=& \frac{1}{2\kappa^2}\left[f_{\cal G}({\cal G}) + 24^2 H^2 \left( 3H^4 + 20 H^2 {\dot H}^2
+ 6 {\dot H}^3 + 4H^3 \ddot H + H^2 \dddot H \right) f_{\cal G}''({\cal G}) \right. \nn
&& \left. - 24^3 H^5 \left(2{\dot H}^2 + H \ddot H + 4 H^2 \dot H \right)^2 f_{\cal G}'''({\cal G})\right]
+ p_\mathrm{matter}\ .
\eea
In the same way one can get the effective gravitational pressure and energy density so that the
equations of motion for arbitrary modified gravity can be rewritten in the universal FRW form typical
for General Relativity:
\be
\label{IV}
\frac{3}{\kappa^2} H^2 = \rho_\mathrm{eff}\, , \quad
p_\mathrm{eff} = - \frac{1}{\kappa^2} \left( 2\dot H + 3 H^2 \right)\, .
\ee
There are just standard FRW gravitational equations.

Formally, the modified gravity equations (\ref{IV}) have just the same form as for GR with matter.
The only difference is that the effective energy-density and pressure are
caused by
extra gravitational terms due to the modification of the GR Lagrangian. It is quite well-known
that equations of motion (\ref{IV}) can describe the accelerating (early-time or late-time)
 epoch when all or part of the (effective) energy conditions are violated.
For instance, when $w_\mathrm{eff} < -1$, one gets the effective phantom (super)acceleration
evolutionary phase (all energy conditions are violated).
When $-1<w_\mathrm{eff}< - 1/3$ ($\rho_\mathrm{eff} = w_\mathrm{eff} \rho_\mathrm{eff}$),
one gets the effective quintessence acceleration epoch, in this case,
only part of energy conditions is violated.
For $w_\mathrm{eff} = -1$, the de Sitter acceleration evolution emerges.

The remarkable property of phantom/quintessence accelerating universe is the occurrence of the
finite-time future singularity which is caused by the violation of the energy conditions.
The phantom evolution when it is not transient always ends up with so-called Big Rip singularity.
The quintessence accelerating evolution may enter (or may not enter) to
soft finite-time
future singularities. The classification of four possible finite-time future
singularities for late-time accelerating universe has being made in ref.
\cite{Nojiri:2005sx}.

The clarifying remark is in order.
Big number of phantom/quintessence dark energy models (say, fluid, scalar,
spinor, DBI, tachyon, etc.)
brings the evolution to the finite-time future singularity. From the
above, equivalent description it is clear
that number of modified gravities also describes the effective
phantom/quintessence accelerating
universe which evolves to finite-time future singularity.
{\it It is important to understand that the singularity occurrence is general feature
of the phantom/quintessence dark energy models which may be fluid, scalar, or modified gravity.}
Moreover, even the classification of finite-time singularities for modified gravity is
given by the same scheme \cite{Nojiri:2005sx} where $p$ and $\rho$ should be substituted by
$p_\mathrm{eff}$, $\rho_\mathrm{eff}$:
\begin{itemize}
\item Type I (``Big Rip'') : For $t \to t_s$, $a \to \infty$,
$\rho_\mathrm{eff} \to \infty$ and $\left|p_\mathrm{eff}\right| \to \infty$.
This also includes the case of $\rho_\mathrm{eff}$, $p_\mathrm{eff}$ being finite at $t_s$.
\item Type II (``sudden'') : For $t \to t_s$, $a \to a_s$,
$\rho_\mathrm{eff} \to \rho_s$ and $\left|p_\mathrm{eff}\right| \to \infty$
\item Type III : For $t \to t_s$, $a \to a_s$,
$\rho_\mathrm{eff} \to \infty$ and $\left|p_\mathrm{eff}\right| \to \infty$
\item Type IV : For $t \to t_s$, $a \to a_s$,
$\rho_\mathrm{eff} \to 0$, $\left|p_\mathrm{eff}\right| \to 0$ and higher derivatives of $H$ diverge.
This also includes the case in which $p_\mathrm{eff}$ ($\rho_\mathrm{eff}$)
or both of $p_\mathrm{eff}$ and $\rho_\mathrm{eff}$
tend to some finite values, while higher derivatives of $H$ diverge.
\end{itemize}

Indeed, it has been observed the approach to Big Rip singularity in the modified gravity in
ref.\cite{Abdalla:2004sw}. However, it was demonstrated there that phantom phase becomes transient
due to presence of $R^2$-term and no singularity occurs.
In ref.\cite{Briscese:2006xu}, it was clearly demonstrated the appearance of Big Rip type
singularity in several $f(R)$ modified gravity models (see also examples
of modified gravity which describe the phantom era and occurrence of Big Rip singularity
in \cite{Nojiri:2006ri}).

The explicit construction of $f(R)$ gravity models which lead to all four possible types of
future singularity after the corresponding quintessence/phantom era has
been presented in ref.\cite{Nojiri:2008fk}.
Even more explicit models of $f(R)$ and other modified gravities which realize
the four possible finite-time future singularity have been presented in ref.\cite{Bamba:2008ut}.
The search of singular modified gravity models was done using the reconstruction scheme developed
in refs.\cite{reconstruction}. In this way, the modified gravity which
realizes finite-time Type I,II,III or IV future singularity may be
constructed in the same way as for traditional DE models (scalar,
vector, spinor, etc).
Moreover, it was shown in
refs.\cite{Nojiri:2008fk,Bamba:2008ut,Capozziello:2009hc}, how
to make the non-singular model from the initially singular modified gravity.
Explicitly it was considered the addition of $R^2$-term (as in the
non-singular
modified gravity model \cite{Abdalla:2004sw}), or some other gravitational terms \cite{Bamba:2008ut}
relevant at the early universe, or the account of the quantum
contribution due to conformal
anomaly as scenario for curing the future singularity of any type.
(For related discussion of Type II finite-time future singularity in specific model of
$f(R)$ gravity and its avoidance by the same trick \cite{Abdalla:2004sw} of  adding $R^2$-term, see
refs.\cite{sami}).
It is interesting to note that above four types of future singularities in FRW space-time
reappear in spherically symmetric space-time as energy density and pressure singularities
at finite radius \cite{Nojiri:2009uu}. It is precisely these singularities
which may lead to problems in the formation of relativistic stars and
black holes in any dark energy model evolving to future singularity.
In the next section, we discuss the realistic $f(R)$ gravity which is non-singular and which
unifies the early-time inflation with late-time acceleration.

\section{Non-singular realistic $f(R)$ model unifying the early-time
inflation with late-time acceleration}

Let us consider the following $f(R)$ model \cite{Nojiri:2007cq}:
\be
\label{V}
f(R) = \frac{\alpha R^{m+l} - \beta R^n}{1 + \gamma R^l}\,.
\ee
For the case $m=l=n$, the model has been analyzed in ref.\cite{Nojiri:2007cq}.
It was shown that the model may describe the realistic unification of
early-time inflation with late-time acceleration (for a first proposal of
the modified gravity
with such unification, see \cite{Nojiri:2003ft}). For the realistic unification,
$\beta\gamma/\alpha \sim 10^{228(n-1)}\,\mathrm{eV}^2$ in the solar system.
This gives extremely small (non-observable) correction to the Newton law if $n\geq 2$.
In this case, there is no the observable fifth force (the corresponding force is
much below the observational bounds).
It is not difficult to check that for $n=2$, the above model (\ref{V}) which passes the
local tests is also non-singular.
Hence, the model (\ref{V}) suggests the generalization of the scenario to cure the
future finite-time singularity via adding the $R^2$-term \cite{Abdalla:2004sw}.
Actually, the same mechanism is used: one can consider the addition of the term $R^2 \tilde f(R)$
where $\lim_{R\to 0} \tilde f(R) = c_1$, $\lim_{R\to \infty} \tilde f(R) = c_2$ as
the scenario to remove the future singularity. In principle, other higher
derivative terms may be also proposed for this purpose.
Hence, one can consider the realistic dark energy alternative $f(R)$ gravity
from the class suggested in ref.\cite{hu}
(or more general realistic models \cite{Nojiri:2007as} unifying the early-time inflation with
late-time acceleration).

In case that such theory contains the finite-time future singularity (as
it happens
in many other dark energy models \cite{Nojiri:2005sx}), one can add
 $R^2$-term \cite{Abdalla:2004sw} or $R^2 \tilde f(R)$ (see above) in order to cure the finite-time
singularity. Of course, one can also take into account quantum effects due
to conformal
anomaly \cite{Nojiri:2005sx,Bamba:2008ut} or quantum gravity effects
\cite{eli} or
generalized fluid of special form \cite{Bamba:2008ut,joseluis} to prevent
the occurrence of
future singularity.
Moreover, the use of the function of the form (\ref{V}) to cure the finite-time singularity
seems to be the most promising one.
The reason is that the terms of such sort are mainly relevant at the early-time era,
so eventually they do not spoil the local tests of the theory. From
another side, such terms give the contribution to the cosmological perturbations so
probably they may be used to make that more consistent with the observational data.

As the realistic example, we consider the viable theory \cite{Nojiri:2007as}:
\be
\label{VI}
f(R) = - \frac{\left(R - R_0\right)^{2k+1} + R_0^{2k+1}}{f_0 + f_1
\left\{\left(R - R_0\right)^{2k+1} + R_0^{2k+1} \right\}}\, .
\ee
It has been shown \cite{Nojiri:2007as} that for $k\geq 10$ such modified gravity passes
the local tests. It also unifies the early-time inflation with dark energy epoch.

In (\ref{VI}), $R_0$ is current curvature $R_0\sim \left(10^{-33}\,\mathrm{eV}\right)^2$.
We also require,
\be
\label{Uf7}
f_0 \sim \frac{R_0^{2n}}{2} \ ,\quad
f_1=\frac{1}{\Lambda_i}\ .
\ee
Here $\Lambda_i$ is the effective cosmological constant in the inflation epoch.
When $R\gg \Lambda_i$, $f(R)$ (\ref{VI}) behaves as
\be
\label{VIII}
f(R) \sim - \frac{1}{f_1} + \frac{f_0}{f_1^2 R^{2n+1}}\ .
\ee
We now use the trace equation, which is the trace part of the FRW
gravitational equation
\be
\label{Scalaron}
3\Box f'(R)= R+2f(R)-Rf'(R)-\kappa^2 T\ .
\ee
Here $T$ is the trace of the matter energy-momentum tensor. Then assuming
the FRW metric with
flat spatial part, one finds
\be
\label{R}
R \sim \left(t_0 - t\right)^{-2/\left(2n+3\right)}\ ,
\ee
which diverges at finite future time $t=t_0$.
By a similar analysis, we can show that if $f(R)$ behaves as $f(R) \sim R^\alpha$ for large
$R$ with a constant $\alpha$, a future singularity appears if $\alpha>2$ or $\alpha<0$.
Conversely if $2\geq \alpha \geq 0$, the singularity does not appear.
Then by adding the previous
term $R^2 \tilde f(R)$, where $\lim_{R\to 0} \tilde f(R) = c_1$,
$\lim_{R\to \infty} \tilde f(R) = c_2$, to $f(R)$ in (\ref{VI}),
the future singularity (\ref{R}) disappears.

Let us observe the above situation in more detail. Now we assume
\be
\label{X}
f(R) \sim F_0 + F_1 R^\alpha\ ,
\ee
when $R$ is large. Here $F_0$ and $F_1$ are constants where $F_0$ may
vanish
but we assume $F_1\neq 0$.
In case of (\ref{VIII}), we have
\be
\label{XI}
F_0 = - \frac{1}{f_1} \ ,\quad
F_1 = \frac{f_0}{f_1^2} \ ,\quad
\alpha = -\left(2n+1\right) \ .
\ee
Under the assumption (\ref{X}), the trace equation (\ref{Scalaron}) gives
\be
\label{XII}
3 F_1 \Box R^{\alpha -1} = \left\{
\begin{array}{ll} R & \ \mbox{when $\alpha<0$ or $\alpha=2$} \\
\left(2-\alpha\right) F_1 R^\alpha & \ \mbox{when $\alpha>1$ or $\alpha\neq 2$}
\end{array} \right. \ .
\ee
In the FRW background with flat spatial part, when the Hubble rate has a
singularity as
\be
\label{XIII}
H \sim \frac{h_0}{\left(t_0 - t\right)^\beta}\ ,
\ee
with constants $h_0$ and $\beta$, the scalar curvature $R=6\dot H + 12 H^2$ behaves as
\be
\label{XIV}
R \sim \left\{ \begin{array}{ll}
\frac{12h_0^2}{\left(t_0 - t\right)^{2\beta}} & \ \mbox{when $\beta>1$} \\
\frac{6 h_0 + 12 h_0^2}{\left(t_0 - t\right)^2} & \ \mbox{when $\beta=1$} \\
\frac{6\beta h_0}{\left(t_0 - t\right)^{\beta + 1}} & \ \mbox{when $\beta<1$}
\end{array} \right. \ .
\ee
In (\ref{XIII}) or (\ref{XIV}), $\beta\geq 1$ case corresponds to Type I (Big Rip) singularity,
$1>\beta>0$ to Type III, $0>\beta>-1$ to Type II, and $\beta<-1$ but $\beta\neq\mbox{integer}$ to
Type IV.
By substituting (\ref{XIV}) into (\ref{XII}), we find that there are two classes of consistent solutions.
The first solution is specified by $\beta =1$ and $\alpha>1$ but
$\alpha\neq 2$ case, which corresponds to the Big Rip
($h_0>0$ and $t<t_0$) or Big Bang ($h_0<0$ and $t>t_0$) singularity at $t=t_0$.
Another one is $\alpha<1$, and
$\beta = - \alpha/\left(\alpha - 2\right)$ ($-1<\beta<1$) case, which
corresponds to (\ref{R}) and to the II Type future singularity.
In fact, we find $\alpha = - 2n -1$ and therefore $\beta + 1 = - 2/(2n+3)$.
We should note that when $\alpha=2$, that is, $f(R) \sim R^2$, there is no any singular solution.
Therefore if we add the above
term $R^2 \tilde f(R)$, where $\lim_{R\to 0} \tilde f(R) = c_1$,
$\lim_{R\to \infty} \tilde f(R) = c_2$, to $f(R)$ in (\ref{VI}), the added term dominates and
modified $f(R)$ behaves as $f(R)\sim R^2$,
the future singularity (\ref{R}) disappears.
We also note that if we add $R^n$-term $n=3,4,5,\cdots$, the singularity becomes (in some sense)
worse since this case corresponds to $\alpha=n>1$, that is Big Rip case. 
Using the potential which appears when we transform $F(R)$-gravity to 
scalar-tensor theory \cite{Nojiri:2008fk}, it has been found that the future singularity 
may not appear in case $0<\alpha<2$. 

Thus, we demonstrated that combination of two realistic modified gravities
where one of them is non-singular one may heal the finite-time future
singularity of combined realistic model. On the same time, the nice
properties of the
models: the successful passing of local tests as well as unified
description of the inflation with dark energy epoch remain to be the same.

\section{Discussion}

In summary, using the fluid representation, we demonstrated that future
singularity appearance is typical property of number of modified gravities
with the effective quintessence/phantom late-time behavior in the same
way as for other simpler (scalar/fluid) dark energy models.
These four different future cosmological singularities may manifest
themselves as effective pressure/energy-density divergence at finite
radius in spherically-symmetric space. This may cause number of problems
with black holes and
relativistic stars formation for {\it any singular dark energy} (as is shown in
\cite{Nojiri:2009uu}), not only for
alternative gravity dark energy models. Moreover, the easiest way to cure
the future singularity is again to call for the modification of gravity,
for instance, adding the $R^2$-term which is relevant at the very early
universe. Hence, starting from singular fluid/scalar dark energy one is
forced to modify its gravitational sector in order to cure the
singularity. Is not then more natural to start from the modified gravity
from the very beginning?
Moreover, the
combination of non-singular realistic modified gravity with singular
realistic model unifying the early-time inflation with late-time
acceleration leads to the resulting realistic non-singular theory. We
concentrated mainly on
$f(R)$ gravity but other modified gravities may show very similar singular
behavior for effective quintessence/phantom late-time era as is explained
in ref.\cite{Bamba:2008ut}. The same universal tools to cure the future
singularity may be used for any modified gravity. Finally, it is interesting 
to mention that above scheme easily accounts for description of DM effects within 
the same modified gravity (for review of DM properties from modified gravity, see \cite{far}).

\begin{theacknowledgments}

The work is supported in part by 
MEC (Spain) project FIS2006-02842 and AGAUR (Catalonia) 2009SGR-994, by
JSPS Visitor Program (Japan) (S.D.O);
and 
Global COE Program of Nagoya University provided by the Japan Society
for the Promotion of Science, G07 (S.N.).

\end{theacknowledgments}

\end{document}